\let\origvec\vec
\documentclass[runningheads]{llncs}

\let\vec\origvec

\usepackage{silence}
\usepackage{url}
\usepackage{amsmath}
\usepackage{caption}
\usepackage{hyperref}
\usepackage{multicol}
\usepackage{graphicx}
\usepackage{enumitem}
\usepackage{mathtools}
\usepackage{adjustbox}
\usepackage[colorinlistoftodos]{todonotes}
\usepackage[ruled,vlined,linesnumbered]{algorithm2e}

\graphicspath{ {images/} }

\AtBeginDocument{
  \providecommand\BibTeX{{
    \normalfont B\kern-0.5em{\scshape i\kern-0.25em b}\kern-0.8em\TeX}}}

\begin{document}
\title{On the Vulnerability of Community Structure in Complex Networks}

\author{Viraj Parimi\inst{1}\thanks{The work was done when Viraj was an undergraduate student at IIIT-Delhi, India.} \and
Arindam Pal\inst{3,4,6} \and
Sushmita Ruj \inst{3,5} \and 
Ponnurangam Kumaraguru \inst{2} \and 
Tanmoy Chakraborty \inst{2}}

\authorrunning{V. Parimi et al.}

\institute{Carnegie Mellon University, Pittsburgh, USA \and
IIIT-Delhi, New Delhi, India \and
Data61, CSIRO, Sydney, Australia \and
Cyber Security CRC, Canberra, Australia \and
Indian Statistical Institute, Kolkata, India \and
University of New South Wales, Sydney, Australia}

\maketitle

\begin{abstract}
In this paper, we study the role of nodes and edges in a complex network in dictating the  robustness of a community structure towards structural perturbations. Specifically, we attempt to identify all vital nodes, which, when removed, would lead to a large change in the underlying community structure of the network. This problem is critical because the community structure of a network allows us to explore deep underlying insights into how the   function and topology of the network affect each other. Moreover, it even provides a way to condense large networks into smaller modules where each community acts as a meta node and aids in more straightforward network analysis. If the community structure were to be compromised by either accidental or intentional perturbations to the network, that would make such analysis difficult. Since identifying such vital nodes is computationally intractable, we propose a suite of heuristics that allow to find solutions close to the optimality. To show the effectiveness of our approach, we first test these heuristics on small networks and then move to more extensive networks to show that we achieve similar results. Further analysis reveals that the proposed approaches are useful to analyze the vulnerability of communities in networks irrespective of their size and scale. Additionally, we show the performance through an extrinsic evaluation framework -- we employ two tasks, i.e., link prediction and information diffusion, and show that the effect of our algorithms on these tasks is higher than the other baselines.
\end{abstract}

\section{Introduction}
\label{introduction}

A large body of research in complex networks involves the study and effects of community structure as it is one of the salient structural characteristics of real-world networks. A network is said to have a community structure if it can be grouped easily into sets of nodes. Each set of nodes is densely connected internally and sparsely linked externally. Research in this field is broadly classified into two categories -- first, where one detects the community structure within a given network and the other where one studies the properties of a community structure to infer more details about the network. A variety of methods have been proposed that target the former issue as described  \cite{Lancichinetti,chakraborty2017metrics}. The advantage of such algorithms is that it provides us with an efficient and approximate clustering of nodes that allows us to condense large networks to smaller ones owing to their mesoscopic structure. Within the second paradigm, the ability to detect vital nodes is of significant practical importance. It provides insight into how a network functions and how the network topology change affects the interactions between the nodes within the network. Exploring this structural vulnerability of the network allows us to prepare beforehand if the network is affected by undesired perturbations and adversarial attacks. A significant factor in understanding this is to analyze the network and comprehend the effect of these vital nodes' failure on the community structure of the network. 

In this paper, we attempt to identify and investigate some vital nodes in a network,  whose removal highly affects the network's community structure. Formally, given a network $G(V, E)$ and a positive integer $k$, we intend to find a set $S \in V$ consisting of $k$ nodes whose removal leads to the maximum damage of the community structure. The change in the community structure is quantified using different measures such as Modularity \cite{Modularity}, Normalized Mutual Information \cite{NormalizedMutualInformation}, Adjusted Rand Index \cite{AdjustedRandIndex}, etc.

There are many real-world applications of this problem. Consider a power grid network where a power outage is a frequently occurring event. Most power networks have a regional hub that caters to the needs of nearby power stations. In such a scenario, the vendor of this power grid needs to make quick decisions about how the failure of some nodes in the network would affect the customers. The solution would be to ensure that crucial nodes in this network have enough backup so that the restoration process can move effortlessly. Another application would be the railway networks, where inadvertent cutting of routes to certain stations can cause significant problems for the city residents. Hence, the government needs to ensure that routes to certain critical stations have redundancies so that if one route gets cut off, then the trains can utilize other routes. This problem also has applications in Online Social Networks such as the worm containment problem \cite{WormContainmentProblem}. This knowledge would provide helpful insights into protecting sensitive nodes once worms spread out into the network. In all of the issues mentioned above, it is evident that one needs to study the structural integrity of the communities underlying in the network. Note that a minor structural change that can be as small as removing a node in the network can lead to the community's breakdown that the node was a part of, given that the removed node had a considerable influence on the network. If the removed node were of less significance, that would have less impact on the network's community structure.

Additionally, understanding the network vulnerability from the standpoint of the community structure is essential in real-world settings. The networks that are dealt with here have tremendous size, which adds to the computational overhead and, most importantly, shed light on some latent characteristics shared by different nodes. Since communities can act as meta-nodes, they allow for a more comfortable study of large networks. This reduces the computational overhead and provides useful insights based on the properties shared by the community's nodes that can be exploited to understand the network's vulnerability.

We propose a hierarchical greedy approach that selects communities based on the community-centric properties in phase 1 and then, within that community, selects the most vulnerable nodes in phase 2. We test this algorithm on six real-world datasets of varying sizes. Our empirical results indicate that the algorithm can identify properties that contribute most towards community structures' vulnerabilities in a network. The past work in this domain \cite{Sankaran,Alim} is restricted to smaller networks, but our work extends the scope towards even more extensive networks with the number of nodes in the order of millions. 

In summary, our contributions in this paper are as follows:
\begin{itemize}
\item We study the structural vulnerability of communities in networks and assess the impact of nodes' failure on the underlying community structure.
\item We suggest few heuristics, including a hierarchical greedy approach that allows for identifying such critical nodes in the network that profoundly impact the community structure.
\item We conduct experiments on real-world datasets and show the effectiveness of the heuristics that we propose.
\item We propose a novel task-based strategy to evaluate the extent of correctness of the algorithm extrinsically. This allows us to estimate the performance of our algorithm in a real-world context.  
\end{itemize}

The remaining part of this paper is as follows. We discuss the literature review on community detection and vulnerability assessment in Section \ref{related_work}. We formalize our problem in Section \ref{problem_statement}. We then discuss some preliminaries in Section \ref{preliminaries}. We present our proposed methodology in Section \ref{methodology}.  Section \ref{datasets} describes the datasets used to evaluate the proposed approach. In Section \ref{experiments} we provide the results of our method when applied to these datasets and briefly discuss the evaluation strategy how we go about validating our proposed method on larger datasets. We put forward our conclusion in Section \ref{conclusion}.

\section{Related Work}
\label{related_work}

This section first presents the literature on community detection algorithms and then discusses community vulnerability analysis.

\subsection{Community Detection}

Community detection, a task of grouping similar nodes together, is a significant problem. A lot of work has been done in the past to come up with a solution effectively. Numerous approaches have been developed and applied to detect community structure. For instance, a hierarchical agglomerative algorithm was proposed by Newman et al.\cite{Football}. An extensive literature survey can be found in \cite{Lancichinetti}. Here we briefly mention some of the popular approaches. 

Initial efforts at community detection assumed that the nodes are densely connected within a community and sparsely connected across communities. Under this assumption, the algorithms proposed were targeted towards community detection in static networks. Such efforts involved several approaches such as modularity optimization \cite{Louvain,Clauset2004,Guimera,Modularity,newman03fast}, clique percolation \cite{Farkas_2007,PalEtAl05}, information-theoretic approaches \cite{Rosvall7327,Rosvall1118}, and label propagation \cite{LabelPropagation1,JieruiXie,Xie}. Furthermore, spectral partitioning \cite{Newman_2013,richardson_spectral_2009}, local expansion \cite{jeffrey_baumes,Lancichinetti_2009}, random-walk based approaches \cite{DeMeo,rushed_kanawati}, diffusion-based approaches \cite{LabelPropagation1} and significance-based approaches \cite{Lancichinetti_plos} were explored to help in identifying the community instances within a network. Several pre-processing methods \cite{dimacs,Pre-processing_1} were also introduced to improve upon these algorithms. Such methods involved generating a preliminary community structure on a set of nodes and modifying iteratively until all the nodes are covered. Apart from these, several other algorithms were proposed to detect communities in dynamically evolving networks \cite{agarwal2018dyperm,rossetti2018community}. 

Another set of community detection algorithms allows a vertex to be part of multiple communities simultaneously. Such overlapping community detection algorithms used ideas based on local expansion and optimization. These include RankRemoval \cite{RankRemoval} which uses local density function, LFM \cite{Lancichinetti_2009}, and MONC \cite{Havemann_2011} which iteratively maximize a fitness function, and GCE \cite{GCE} which makes use of an agglomerative pipeline to detect overlapping community instances. Other approaches also looked into the idea of partitioning links instead of nodes to discover the network's underlying community structure. The clique percolation method was also explored in CFinder \cite{CFinder}, but since many real-world networks are sparse, these methods generally produced low-quality outputs. Recently, several new ideas were presented, such as \cite{Nepusz2008FuzzyCA} which solved a constrained optimization problem using simulated annealing techniques, and \cite{mix1,mix2,mix3,mix4} which used mixture models to solve the problem. Even a game-theoretic approach \cite{Chen_Nash} was proposed in which a community is equated to a Nash local equilibrium. Non-negative Matrix Factorization \cite{Yang_NMF,Zhang_NMF} framework has also been utilized to identify fuzzy or overlapping community structures. Chakraborty et al. proposed MaxPerm and GenPerm, two greedy approaches which maximize a node-centric metric, called "permanence" to detect disjoint \cite{Football} and overlapping communities \cite{chakraborty2016genperm}. They also proposed a post-processing technique based on permanence to detect overlaps from a disjoint community structure \cite{chakraborty2015leveraging}. Several ensemble-based approaches were also proposed by leveraging the output of disjoint community detection methods \cite{chakraborty2017ensemble,chakraborty2019ensemble,chakraborty2016ensemble}.

\subsection{Community Vulnerability Analysis}

Assessing the structural network vulnerability has received increasing attention. For example, Nguyen et al. \cite{Thai} have proposed a Community Vulnerability Assessment (CVA) problem and suggested multiple heuristic-based algorithms based on the modularity measure of communities in the network. These approaches are restricted to the scope of online social networks and do not cater to general network structures. Another work by Nguyen et al. \cite{Nguyen}  explored the number of \textit{connected triplets} in a network as they capture the strong connection of communities in social networks. They proposed an efficient approximation algorithm to identify triangle breaking points like nodes or links within a network. 

Additionally, different measures and metrics have been proposed to measure the robustness of a network. Such efforts include the average size of a cluster, relative size of the largest components, diameter, and network connectivity. One approach dealt with this problem using the weighted count of loops in a network. Chan et al. \cite{Chan} addressed this problem in both deterministic and probabilistic settings where they suggested solutions based on minimum node cutset. Frank et al. \cite{Frank} outlined a solution that uses the second smallest eigenvalue of a Laplacian matrix of a network and termed it as the algebraic connectivity of that network. Fiedler \cite{Fiedler} proposed four basic attack strategies, namely, ID removal, IB removal, RD removal, and RB removal. ID and RD removal deal with the degree distribution of the network. The only difference is that the second approach changes the removal strategy based on the degree distribution change. IB and RB removal are also similar constructs, but they are based on the betweenness distribution. Holme et al. \cite{Holme} used an algorithm adapted from Google's PageRank providing a sequence of losses that add to the collapse of the network. Allesina et al. \cite{Allesina} evaluated the network characteristics like cyclomatic number and gamma index. They mentioned that such global graph-theoretic indices are not sufficient to measure a network's vulnerability, but they showcase the hierarchy of nodes in the system.

Ramirez et al. \cite{Ramirez} proposed an approach where the community structure's resilience is quantified by introducing disruption in the original network and measuring the change in the community structure temporally, i.e., after the disconnection and during the restoration process. Geubesic et al. \cite{Grubesic} provided a review of various approaches that use the \textit{facility importance} concept to understand the system-wide vulnerability. These concepts include alpha index, beta index etc. They concluded that simple graph-theoretic measures were not sufficient to measure the vulnerability of a network. It also required many local efforts, such as the degree of node. They mentioned that global indicators measure network accessibility, path availability, and local measures to provide better information about node criticality. Sankaran et al. \cite{Sankaran} proposed a new vulnerability metric where they considered a combination of external and internal factors such as connection density. They proposed a non-linear weighted function to combine these factors. However, the proposed method was not proved feasible in practice as the weights of all the elements were assumed to be equal and not self-adjusting to the network. These methods allow us to quantify a community's vulnerability but do not provide us with a set of nodes that contribute to the community's vulnerability. 

The information of critical nodes that contribute to the communities' vulnerability would provide far more insights than just discovering the vulnerable community. As a result, a more comprehensive study is required to assess the vulnerability of general network structures. 

\section{Problem Statement}
\label{problem_statement}

Let $G(V,E)$ be an input graph and let $k$ be the number of nodes that we want to select. Let $\mathcal{A}$ be a community detection algorithm. For a vertex set $S \in V$, let $G[S]$ be the subnetwork induced by $S$ and $f(\mathcal{A}(G[V]), \mathcal{A}(G[V \setminus S]))$ is a value function that computes some measure of the difference between the community structures of $G[V]$ and $G[V \setminus S]$ obtained from $\mathcal{A}$. We need to identify a set $S \in V$ of size $k$ which,
\begin{equation}
	\textbf{maximize} \quad f(\mathcal{A}(G[V]), \mathcal{A}(G[V \setminus S]))
\end{equation}

This problem is computationally intractable as shown by Alim et al.  \cite{Thai} and hence requires the use of greedy heuristics to approach the optimal answer.

\section{Preliminaries}
\label{preliminaries}

We used the Louvain algorithm for detecting the underlying community structure. It is a greedy optimization algorithm proposed by Blondel et al. \cite{Louvain}, that tries to optimize the modularity metric of a network and extracts communities from large networks using heuristics. This approach, however, can easily be modified to use with other community detection algorithms as well. 

To quantify the difference between the community structures of $G[V]$ and $G[V \setminus S]$, we use the following measures:

$\bullet$ \textbf{Modularity:} It is a measure to quantify the strength of the division of the network into communities. Networks with high modularity have denser connections within a community and sparse connections across communities. It is defined as follows,
\begin{equation}
    Q = \frac{1}{(2m)}\sum_{vw} \left[ A_{vw} - \frac{k_v 	k_w}{(2m)} \right] \delta(c_{v}, c_{w}),
\end{equation}
    where $m$ = number of edges, $A$ = adjacency matrix, $k_v=$ degree of node $v$, $c_v$ = community label of node $v$, $\delta(c_v, c_w)$ = 1 if $c_i$ = $c_j$ and 0 otherwise.

$\bullet$ \textbf{Normalized Mutual Information:}
It is a measure that quantifies the similarity between two community structures. It produces 1 if two community structures are exactly the same and 0 otherwise. It is defined as follows.
\begin{equation}
    N = \frac{2 \sum\limits_{i = 1}^{c_X} \sum\limits_{j = 1}^{c_Y}\left[n_{ij} \log\left(\frac{n_{ij} n}{x_i y_j}\right)\right]}{(n-k)H(X) + \bar{y}\log(n-k) - \sum\limits_{j = 1}^{c_Y}(y_j \log(y_j))},
\end{equation}
    where $c_X$ = number of communities in community structure $X$, $c_Y$ = number of communities in community structure $Y$, $n_{ij}$ = $\left|X_i \cap Y_j\right|$, $n$ = number of nodes in the network, $x_i$ = $\left|X_i\right|$, $y_i$ = $\left|Y_i\right|$, $\bar{y}$ = total size of communities in $Y$,
    $H(X)$ = entropy of $X$.

$\bullet$ \textbf{Adjusted Rand Index:}
It is another measure of similarity between two data clusterings. It represents the frequency of occurrence of agreements over the total pairs. Its maximum value is 1 which indicates perfect similarity between two clusterings.  It is defined as follows,
\begin{equation}
    R = \frac{ \sum_{ij} \binom{n_{ij}}{2} - \left[\sum_i \binom{a_i}{2} \sum_j \binom{b_j}{2}\right] / \binom{n}{2} }{ \frac{1}{2} \left[\sum_i \binom{a_i}{2} + \sum_j \binom{b_j}{2}\right] - \left[\sum_i \binom{a_i}{2} \sum_j \binom{b_j}{2}\right] / \binom{n}{2} }
\end{equation}
    where $L$ = contingency Table, $n_{ij}$ = $L[i][j]$, $a_i$ = sum  of entries in $i^{th}$ row in $L$, $b_i$ = Sum of entries  in $i^{th}$ column in $L$, $n$ = number of nodes in the network.
    
\section{Proposed Methodology}
\label{methodology}

Given the computation intractability of the problem statement, we first chunk our approach into two sections. We analyze the structural properties of a small network and generate ground-truth data. This data provides us a way to compare our proposed heuristics, thereby quantifying the effectiveness of these heuristics. 

\begin{algorithm}
\DontPrintSemicolon
\SetAlgoLined
\SetKwInOut{Input}{Input}\SetKwInOut{Output}{Output}
\Input{Network $G = (V, E)$, $k$, a community detection algorithm $\mathcal{A}$, a value function $F$}
\Output{Set of nodes whose size is $k$}
\BlankLine
$X \gets$ Run community detection algorithm $\mathcal{A}$ on $G$.\;
$C \gets$ Generate all the combination of nodes in $V$ of size $k$.\;
\ForEach{$C_i \in C$} {
    $G' \gets$ Remove $C_i$ from $G$.\;
    $Y \gets$ Run community detection algorithm $\mathcal{A}$ on $G'$.\;
    Compute $F$ by comparing $Y$ and $X$.\;
    Return a $C_i$ which maximizes $F$.\;
}
\caption{Exhaustive Algorithm}
\label{algo_1}
\end{algorithm}

The thorough approach to gather this information is described in Algorithm \hyperref[algo_1]{1}. This approach compares the networks' community structures before and after structural perturbations, where similarity scores for each combination of nodes are computed.

Yang et al. provide a comparative analysis of significant community detection algorithms, including Edge Betweenness, Fastgreedy, and Infomap. In Algorithm \hyperref[algo_1]{1}, we generate all possible combinations of nodes of size $k$ and then analyze the effect of each such combination to see which minimized the target value function more. Let's consider the computational complexity of the community detection algorithm to be $D$. This means that the complexity of this algorithm is $O(\max(C(n,k), D))$ where $n = |V|$ and $k$ is the budget. Note that $D$ is generally defined in terms of $|V|$ and $|E|$, so this term becomes more dominant for larger networks, thereby increasing the algorithm's computational complexity.

\begin{algorithm}
\DontPrintSemicolon
\SetAlgoLined
\SetKwInOut{Input}{Input}\SetKwInOut{Output}{Output}
\Input{Network $G = (V, E)$, $k$, a structural property to rank the nodes $P$, a community detection algorithm $\mathcal{A}$, a value function $F$}
\Output{Set of nodes whose size is $k$, score}
\BlankLine
$X \gets$ Run community detection algorithm $\mathcal{A}$ on $G$. \;
$R \gets$ Rank the nodes in $G$ based on the structural property $P$. \;
$G' \gets$ Remove top $k$ nodes from $G$ based on $R$. \;
$Y \gets$ Run community detection algorithm $\mathcal{A}$ on $G'$. \;
Compute the value function $F$ by comparing $X$ and $Y$. \;
Return the set of top $k$ nodes in $R$ along with the score of the value function $F$. \;
\caption{Network Based Greedy Approach}
\label{algo_2}
\end{algorithm}

Next, we propose a naive network-based greedy approach defined in Algorithm \hyperref[algo_2]{2}. This algorithm takes in a property as an input and ranks the nodes in the input network based on the property specified. It greedily removes the top $k$ nodes based on their ranks and then evaluates the underlying community structure using a community detection algorithm. The output of this algorithm computes the value function and returns the set of nodes removed along with the evaluated value function score. The structural properties which were used are as follows,
  \begin{itemize}
    \item {\bf Clustering Coefficient} - We use the global coefficient, which is defined as the number of closed triplets over the total number of triplets where a triplet is a set of three nodes that are connected by either two or three undirected edges. The complexity of calculating this property for a node is $O(|V|^3)$. 
    \item {\bf Degree Centrality} - It is defined as the number of edges that are incident upon a node. The time complexity of this metric is $O(|V| + |E|)$.
    \item {\bf Betweenness Centrality} - We use the betweenness centrality estimate defined by Freeman. \cite{BetweenessCentrality} as the number of times a node acts as a bridge along a shortest path route between two other nodes. Time complexity of this metric is $O(|V||E| + |V|^2)$.
    \item {\bf Eigenvector Centrality} - It is a measure of the influence of a particular node in the network \cite{EigenvectorCentrality}. This centrality estimate is based on the intuition that a node is more central when there are more connections within its local network. The time complexity of calculating this metric for a node is $O(|V|^3)$.
    \item {\bf Closeness Centrality} - It measures how easily other vertices can be reached from a particular vertex \cite{ClosenessCentrality1,ClosenessCentrality2}. Time complexity of this metric is $O(|V||E| + |V|^2)$.
   	\item {\bf Coreness} - The coreness of a node is $k$ if it is a member of a $k$-core but not a member of a $k+1$-core where a $k$-core  is a maximal subnetwork in which each vertex has at least degree $k$  \cite{Coreness}. The time complexity of this metric is $O(|E|)$.
    \item {\bf Diversity} - The diversity index of a vertex is estimated by the normalized Shannon entropy of the weights of the edges incident on a vertex \cite{Diversity}. The time complexity of calculating this metric is $O(|V| + |E|)$.
    \item {\bf Eccentricity} - It is defined as the shortest maximum distance from the vertex to all the other vertices in a network. The time complexity of this metric is $O(|V|^2 + |V||E|)$.
    \item {\bf Constraint} - Introduced by Burt \cite{Constraint}, this measure estimates the time and energy that are concentrated on a single cluster. This measure would be higher for a node that belongs to a small network, and also, all the contacts are highly connected. The time complexity of calculating this metric is $O(|V|+|E|+|V|d^2)$ where $d$ is the average degree. 
    \item {\bf Closeness Vitality} - It is defined as the change in the distance between all node pairs when the node in focus is removed. It is based on the Wiener Index, which is defined as the sum of distances between all node pairs \cite{ClosenessVitality}. The metric's time complexity is $O(|E|\log|V|)$.
   \end{itemize}

This algorithm takes in a community detection algorithm and a structural property of a node as inputs. If the computational complexity of the community detection algorithm is $D$ and for calculating the structural property for all nodes is $S$ as discussed above, then the total computational complexity of Algorithm \hyperref[algo_2]{2} is $O(\max(S, D))$.

\begin{algorithm}[!t]
\DontPrintSemicolon
\SetAlgoLined
\SetKwInOut{Input}{Input}\SetKwInOut{Output}{Output}
\Input{Network $G = (V, E)$, $k$, a global community centric property $P_c$, a node centric property $P_n$, a community detection algorithm $\mathcal{A}$, a value function $F$}
\Output{Set of nodes whose size if $k$, score}
\BlankLine 
\SetKwFunction{FMain}{best\_community}
\SetKwProg{Fn}{Function}{:}{}
\Fn{\FMain{network $G$, community structure $X$}} {
	\ForEach{$X_i \in X$} {
    	$G' \gets$ Create a subnetwork from $G$ with only the vertices from $X_i$. \;
        $R_g \gets$ Rank each such $G'$ based on the community-centric property $P_c$.  \;
        Return $X_i$ whose induced subnetwork $G'$ ranked above others based on $R_g$. \;
    }
}
\SetKwFunction{FMain}{best\_node}
\SetKwProg{Fn}{Function}{:}{}
\Fn{\FMain{network $G$, community structure $X$}} {
    $G' \gets$ Create the subnetwork $G'$ from $G$ which is induced from $X$. \;
	$R_n \gets$ Rank the nodes in $G'$ based on the node centric property $P_n$. \;
    Return the top node that ranks above others based on $R_n$. \;
}
$X \gets$ Run community detection algorithm $\mathcal{A}$ on $G$ \;
$Y \gets$ Run community detection algorithm $\mathcal{A}$ on $G$ \;
\While{$k$ \text{nodes are not selected} } {
	$X' \gets $ best\_community($G$, $Y$) \;
  	$node \gets $ best\_node($G$, $X'$) \;
    $G' \gets$ Remove $node$ from $G$ and add this node to the output set \;
  	$Y \gets $ Run community detection algorithm $\mathcal{A}$ on $G'$ \;
}
Compute the value function $F$ by comparing $Y$ and $X$. \;
Return the output set of nodes and the score evaluated by the value function $F$. \;
\caption{Community Based Greedy Approach}
\label{algo_3}
\end{algorithm}

The downside of this algorithm is that it does not consider the nodes' effect on the community structure of the networks itself. This is addressed in  Algorithm \hyperref[algo_3]{3}, which also considers the underlying community structure. Here, we propose a hierarchical approach where we choose a community based on some community-centric metric in the first phase. Then in the second phase, we select a node greedily based on its node-centric properties.  The community-centric properties used are as follows:

\begin{itemize}
  \item {\bf Link density}: $D(G) = \frac{2E}{V(V-1)}$, where $E$ is the number of edges in the network and $V$ is the number of vertices in the network.
    \item {\bf Conductance}: Given a graph $G(V,E)$, $\lambda(G) = \frac{s}{v}$, where $s$ is number of vertices with one endpoint in $G$ and another in $\bar{G}$, $v$ is the sum of degree of nodes in $G$. This measure calculates how well-knit a graph is.
    \item {\bf Compactness}: $C(G)$ is defined as the average shortest path lengths within the network $G$.
\end{itemize}

This algorithm takes a community detection algorithm, a node's structural property, and a community-centric property as inputs. The computational complexity of calculating the community-centric property will be constant in time as they are defined in terms of fixed formulas; hence, this would not contribute much to this algorithm's overall complexity. If the computational complexity of the community detection algorithm is $D$ and for calculating the structural property for all nodes is $S$ as discussed above, then the total computational complexity of Algorithm \hyperref[algo_2]{2} is $O(\max(S, D))$.

The algorithms proposed above are sufficient for smaller networks. We can evaluate them with Algorithm \hyperref[algo_1]{1}; however, real-world networks exhibit a much more extensive and complex structure. 

The reason is the inefficiency of Algorithm \hyperref[algo_1]{1} as it is a brute force method. This won't allow for the extraction of the ground truth, which we use to estimate the performance of Algorithm \hyperref[algo_2]{2} and Algorithm \hyperref[algo_3]{3}. 
To counter this, we propose a new task-based approach. Here, the intuition is as follows: if the performance of an extrinsic task, based on the network structure is  $\phi$, then after removing the nodes based on the outputs of Algorithm \hyperref[algo_2]{2} and Algorithm \hyperref[algo_3]{3}, the task would perform $\chi \le \phi$ on the new network structure,  thereby validating the selection of nodes.

Specifically, suppose that a user wants to select vulnerable nodes in an extensive network such that the resulting value function score is maximized. To do so, a straightforward way is to use Algorithm \hyperref[algo_2]{2} and Algorithm \hyperref[algo_3]{3} to select the nodes whose effectiveness can be validated by the results of Algorithm \hyperref[algo_1]{1}. But since the network is extensive, it is quite evident that it is not feasible to use Algorithm \hyperref[algo_1]{1}. To counter this, one would use Algorithm \hyperref[algo_4]{4} to validate the results based on the network's performance drop on the tasks. Since we are using the same algorithms used for small networks, it is evident that the actual problem at hand of maximizing the value function is still of prime focus. Only the way to validate those same results has been changed for more extensive networks.

\begin{algorithm}[!t]
\DontPrintSemicolon
\SetAlgoLined
\SetKwInOut{Input}{Input}\SetKwInOut{Output}{Output}
\Input{Network $G = (V, E)$, $k$, a task $T$, a community detection algorithm $\mathcal{A}$, a value function $F$}
\Output{Set of nodes whose size if $k$, score}
\BlankLine 
\SetKwFunction{FMain}{compute\_task\_performance}
\SetKwProg{Fn}{Function}{:}{}
\Fn{\FMain{Task $T$, network $G_1$, network $G_2$, community structure of $G_1$ $X$, community structure of $G_2$ $Y$}} {
     \If{$T$ is link prediction} {
        	Create a test and train edge list based on the edge set of $G_1$. \;
        	$G_1' \gets$ Create a subnetwork induced by the training set \;
        	Apply the link prediction task using $X$ to decide on the edge probabilities on $G_1'$ \;
        	Compute the $F1$ score for the predicted edges \;
        	Repeat the same process for network $G_2$\;
        	Compare the $F1$ scores for both the networks \;
        }
    \Else {
        Select a random set of seed nodes that are active by default \;
        With $p_i = 0.7$ and $p_o = 0.3$ apply the information diffusion task on $G_1$ using the independent cascade model for 200 iterations. This will give the number of active nodes at the end of the iterations \;
        Repeat the process with $G_2$ \;
        Compare the number of active nodes at the end for both $G_1$ and $G_2$ \;
    }
}
$X \gets$ Run community detection algorithm $\mathcal{A}$ on $G$ \;
$S \gets$ Output from Algorithm \ref{algo_2} or Algorithm \ref{algo_3} which return the target set of nodes \;
$G' \gets$ Remove nodes in $S$ from $G$ \;
$Y \gets$ Run community detection algorithm $\mathcal{A}$ on $G'$ \;
$score \gets $ compute\_task\_performance($T$, $G$, $G'$, $X$, $Y$) \;
\caption{Task Based Approach}
\label{algo_4}
\end{algorithm}

In Algorithm \hyperref[algo_4]{4}, we consider two different tasks, which are described as follows:
%to quantify the performance of the graph,
\begin{enumerate}
    \item {\bf Link Prediction:}  We predict the likelihood of a future association between two nodes knowing that there is no association between those nodes in the current state of the network. Hence, the problem asks to what extent the evolution of a complex network can be modeled using features intrinsic to the network topology itself. Generally, in literature, people use few metrics to assign probabilities to a set of non-edges in a network such as Within-Inter-Cluster defined by Rebaza et al. \cite{Rebaza}, Modified Common Neighbors and Modified Resource Allocation defined by Soundarajan and Hopcroft \cite{Soundarajan}.
	\item {\bf Information Diffusion}:  It is defined as the process by which a piece of information is spread and reaches individuals through interactions. We empirically study the behavioral characteristics of information diffusion models, specifically IC (Independent Cascade), on different community structures. We incorporate the community information in this task by assigning $p_i$ probability to edges inside a community and $p_o$ probability to edges that connect separate communities. We keep $p_i \ge p_o$ as information is more likely to spread among nodes within the same neighborhood as observed by Lin et al.\cite{InformationDiffusion}.
\end{enumerate}

\section{Datasets}
\label{datasets}

To run our experiments extensively, we select six real-world networks of diverse sizes. The datasets used are as follows:
\begin{enumerate}
    \item {\bf Karate Club:} The data was collected from the members of a karate club \cite{KarateClub1,KarateClub2}. Each node represents a club member, and each undirected edge represents a tie between two members of the club. The network has two communities, one formed by "John A" and another by "Mr Hi".
    \item {\bf Football Network:}  Girvan and Newman \cite{Football} collected this network. It contains American football games between division IA colleges during the regular season Fall of 2000. The nodes represent teams identified by names, and edges represent regular-season games between two teams that they connect. The network has twelve communities where each community is signified by the conferences that each college belongs to.
    
    \item {\bf Indian Railway Network:}  This network was used in \cite{RailwayandCoauthorship}, which consists of nodes that represent stations where two stations are connected by an edge if there exists at least one train route between them such that these stations are scheduled halts. The states act as communities, and hence there are 21 communities.
    
    \item {\bf Co-authorship Network:}  This network was collected by  Chakraborty et al. \cite{RailwayandCoauthorship}. This dataset comprises nodes representing an author, and an undirected edge between two authors is drawn if and only if they were co-authors at least once. Each author is tagged with one research field on which he/she has written most papers on. There are 24 such fields, and they act as communities.
    
    \item {\bf Amazon Product Co-purchasing Network:}  This was collected by crawling the Amazon site \cite{AmazonandLiveJournal}. The nodes represent products, and an undirected edge between two nodes represents a frequently co-purchased product. There are 75,149 communities, and only groups containing more than three users are considered.
    
    \item {\bf Live Journal:}  This is a free online blogging community where users declare friendship with each other \cite{AmazonandLiveJournal}. Therefore, each node is a user, and an edge between two users represents a friendship. Users are allowed to form groups, and such user-defined groups form communities. There are 287,512 communities, and only groups containing more than three users are considered.
\end{enumerate}

% \begin{table}[!t]
\begin{table}
\centering
\captionsetup{justification=centering, font=small, skip=10pt}
\caption{Properties of the real-world networks  used in our experiments. We chose 3 small and 3 large networks to extensively show the effects of each algorithm in terms of efficiently computing the vulnerable communities.}
\begin{adjustbox}{width=0.9\textwidth}
\begin{tabular}{|c|c|c|c|}
\hline
\textbf{Dataset}                     & \textbf{\#Nodes} & \textbf{\#Edges} & \textbf{\#Communities}  \\ \hline
Karate Club Network                  & 34               & 78               & 2                                          \\ \hline
Football Network                     & 115              & 613              & 12                                          \\ \hline
Indian Railway Network               & 301              & 1,224            & 21                                      \\ \hline
Co-authorship Network                & 103,667          & 352,183          & 24                                           \\ \hline
Amazon Product Co-purchasing Network & 334,863          & 925,872          & 75,149                                     \\ \hline
Live Journal Network                 & 3,997,962        & 34,681,189       & 287,512                                    \\ \hline
\end{tabular}
\end{adjustbox}
\label{dataset}
\end{table}

\section{Experiments}
\label{experiments}

We divide this section into three subsections to cover all the value functions discussed in Section \ref{preliminaries}. We first present the results of Algorithm \hyperref[algo_1]{1} for smaller networks, which will be used as a benchmark to compare the results of Algorithm \hyperref[algo_2]{2} and Algorithm \hyperref[algo_3]{3} whose results will follow. Using the inferences from these results, we build on our argument and present the results of Algorithm \hyperref[algo_4]{4} to establish similar results even on more extensive networks.

\subsection{Modularity}

\subsubsection{Exhaustive approach:}

Table \ref{exhaustive_modularity} shows the results of Algorithm \hyperref[algo_1]{1} on three small networks when using the modularity as the target value function. We perform the analysis by fixing $k = 5$ \footnote[1]{We choose the value of $k$ to be five because of the following reason. Since we intend to compare our approach with the ground truth data, we first need to generate this ground truth data. For smaller $k$ values, the number of nodes' combinations is more diminutive and keeps increasing exponentially as we increase the value of $k$. To limit the computational time, we restricted $k$ to be five, and beyond that, the number of combination of nodes was too large. Simultaneously, we did not want to choose a smaller $k$, as removing a smaller number of nodes would not have that much impact on the underlying community structure than removing more nodes.}. For the Karate network, we observe that nodes (0, 1, 3, 5, 6) are the most vulnerable as their removal maximizes the difference of modularity scores between the original and the perturbed networks. Similarly, the most susceptible nodes identified for the other two networks are mentioned in Table \ref{exhaustive_modularity}. 

% \begin{table}[ht]
\begin{table}
\centering
\captionsetup{justification=centering, font=small, skip=10pt}
\caption{Effect of the exhaustive algorithm on the small networks. The nodes here indicate the ID of the most vulnerable points in the network when the modularity is utilized as the value function. Since the networks are smaller in size, the budget $k$ is fixed at 5 which is why the algorithm detects only 5 vulnerable nodes. The corresponding modularity score reported is the maximum across all possible combinations of the nodes.}
\begin{tabular}{|c|c|c|}
\hline
\textbf{Network} & \textbf{Nodes}          & \textbf{Modularity} \\ \hline
Karate           & (0, 1, 3, 5, 6)         & 0.13436             \\ \hline
Football         & (23, 33, 24, 32, 45)    & 0.10492             \\ \hline
Railway          & (105, 76, 203, 123, 97) & 0.14723             \\ \hline
\end{tabular}
\label{exhaustive_modularity}
\end{table}

\subsubsection{Network Based Greedy Approach:}
\begin{figure}
\centering
\includegraphics[width=\linewidth]{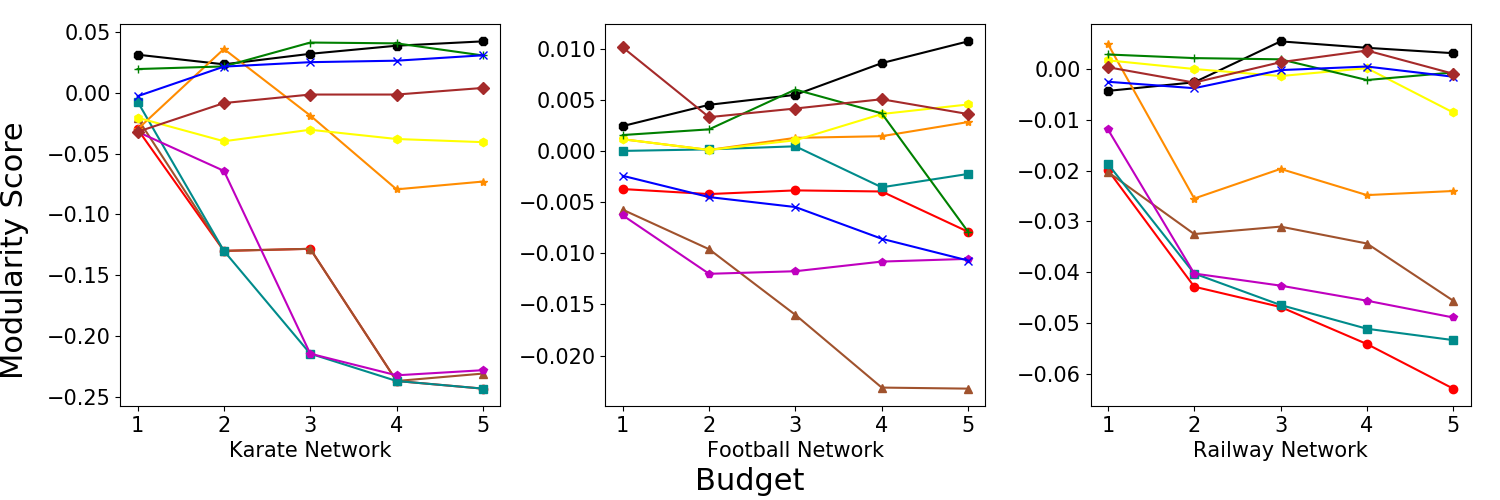}
\vspace{5mm}
\includegraphics[width=\linewidth]{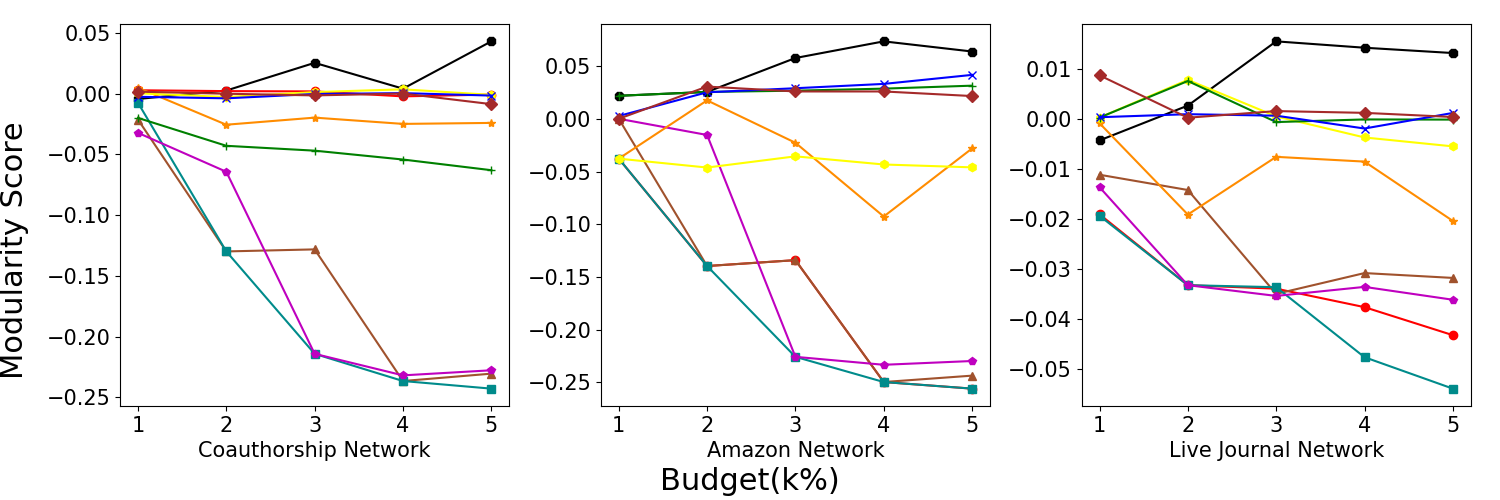}
\vspace{5mm}
\includegraphics[width=\linewidth,height=2cm,keepaspectratio]{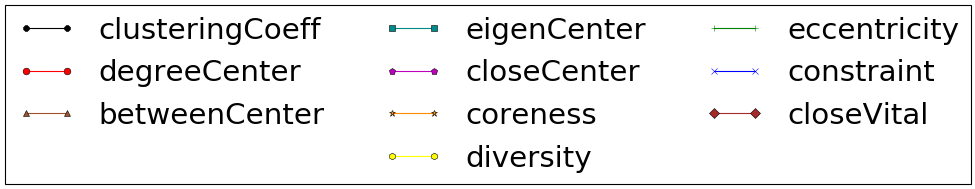}
\captionsetup{justification=centering, font=small}
\caption{Outcome of the network based approach over all the networks with modularity being the target value function. The legend indicates all the structural properties of a network that were used to greedily select nodes. For smaller networks we used $k$ = 5 nodes whereas for larger network s we used 5\% of the nodes in the corresponding network. If we compare the smaller datasets' results with Table \ref{exhaustive_modularity}, we observe that the
maximum values obtained could not attain the optimal values. Across the networks, for higher budget we observe that clustering coefficient turned out to be the best indicator for vulnerability in terms of modularity.}
\label{modularity_network}
\end{figure}

This section presents the analysis results on all the datasets of Algorithm \hyperref[algo_2]{2}. We performed this analysis on all the datasets irrespective of their scale as the algorithm applied was greedy and did not need much time to execute. Moreover, we fix $k = 5$ for smaller networks, but such a removal strategy won't showcase significant effects for more extensive networks. This is because removing just five nodes in more extensive networks won't affect the underlying community structure enough to cause substantial structural perturbations. So, to handle such cases, we instead remove 5\% of the total nodes. From Figure \ref{modularity_network}, we infer that the clustering coefficient as a network-based greedy metric performs better than other greedy metrics when we remove the target five nodes.

Moreover, when we compare the maximum values attained in the smaller networks, we see that this algorithm cannot achieve the optimal answer indicated in Table \ref{exhaustive_modularity}. For example, in the Karate network, the maximum score obtained by Algorithm \hyperref[algo_2]{2} is around 0.05, whereas the optimal answer is 0.13. This indicates that there is a lot of scope for improvement.

% \begin{figure}[!t]

\subsubsection{Community Based Greedy Approach:}

\begin{figure}
\centering
\includegraphics[width=\linewidth]{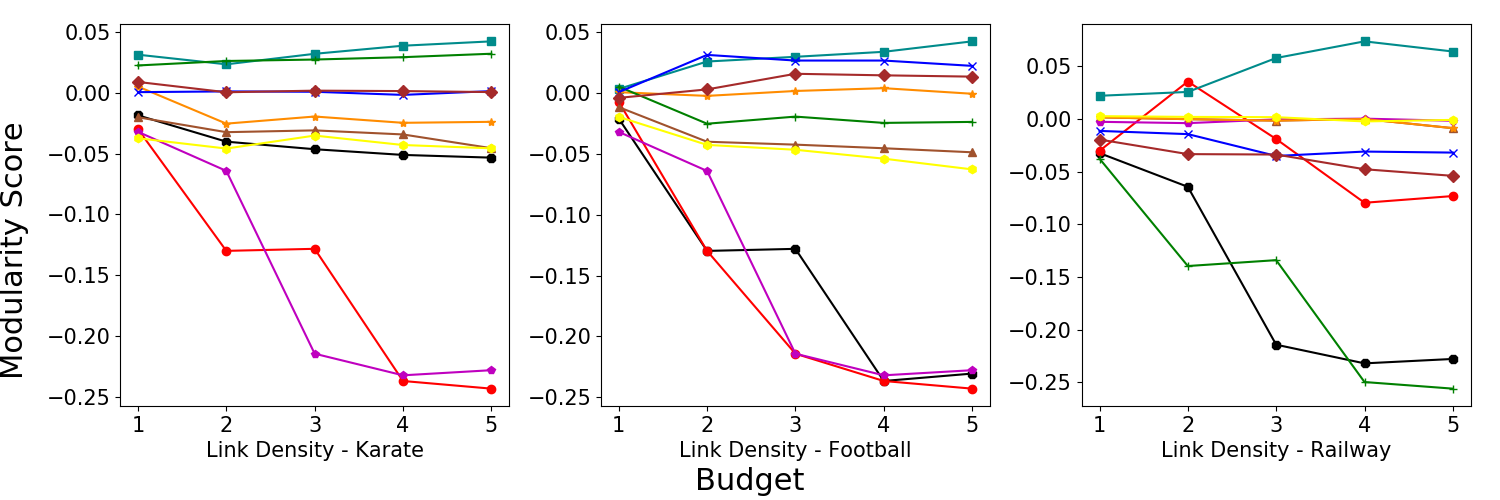}
\includegraphics[width=\linewidth]{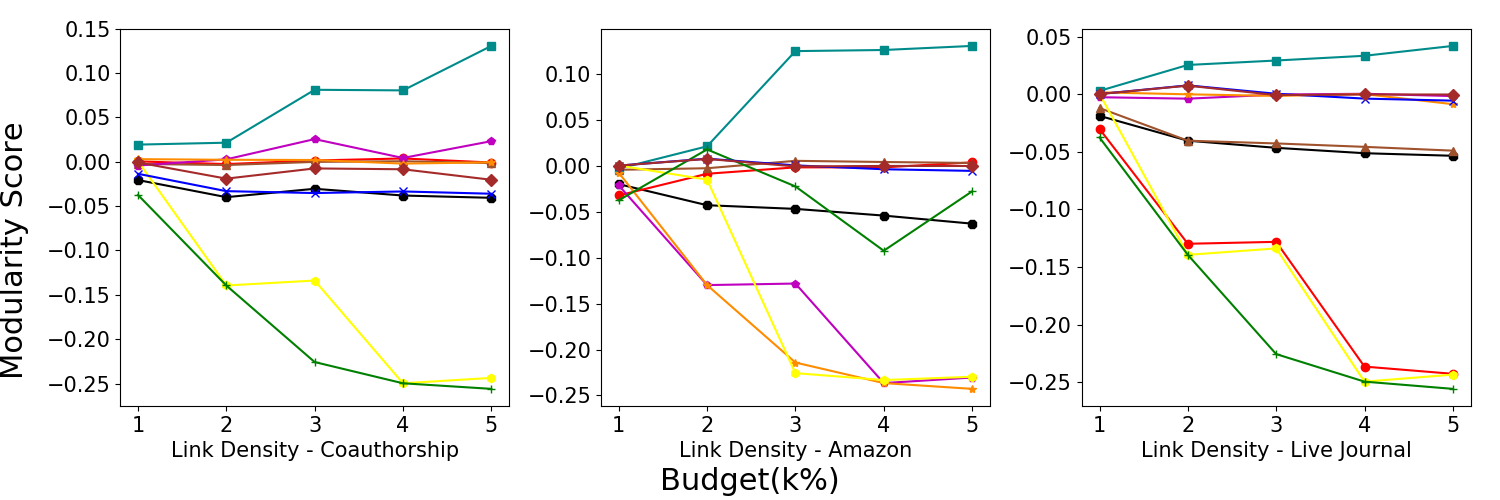}
\includegraphics[width=\linewidth,height=2cm,keepaspectratio]{legend.png}
\captionsetup{justification=centering, font=small}
\captionof{figure}{Results of the community based approach over several datasets with modularity being used as the target value function. These results are reported only for Link Density as it outperformed the other community based greedy metrics as described in Table \ref{modularity_community_compare}. The plots indicate that across the datasets for larger budgets, eigenvector centrality performs better in comparison to other greedy metrics.
The values indicated are close to the ground truth as reported in Table \ref{modularity_network}.}
\label{modularity_community}
\end{figure}

In this section, we evaluate the performance of Algorithm \hyperref[algo_3]{3} over all the datasets. As mentioned previously, we fix $k = 5$ for smaller networks and 5\% for more extensive networks. We compare the different community-centric properties in Table \ref{modularity_community_compare}. Here we present the best modularity scores obtained after applying this algorithm on all the datasets. As this algorithm is also inherently greedy, it also is computationally efficient. Based on Table \ref{modularity_community_compare}, we observe that Link Density performed better than the other community-centric properties as the scores over all the datasets were maximum. Now that we have established that the best community-centric property in a modularity difference maximization setting is link density, we present the node centric properties' results in Figure \ref{modularity_community}. Overall, we ran experiments on the datasets; we found that eigenvector centrality performs better than other greedy metrics.

Additionally, when we compare this algorithm's results with the ground truth data presented in Table \ref{exhaustive_modularity}, we observe that this solution comes close to the optimal solution. For example, in the Railway network, we follow that the best modularity score obtained to be around 0.06, which is close to the ground truth score of 0.14 compared to the 0.01 score obtained from Algorithm \hyperref[algo_1]{1}. So it is evident from this data that the difference between the optimal solution and the current solution has decreased, thereby establishing the superiority of Algorithm \hyperref[algo_3]{3} over \hyperref[algo_2]{2}.

% \begin{table}[!htp]
\begin{table}
\centering
\captionsetup{justification=centering, font=small, skip=10pt}
\caption{Outcome of the community based approach using modularity as the target value function. It shows the effects of different community based metrics have when used to greedily select nodes. Based on this table we observe that Link Density performs better to indicate the vulnerability of nodes in terms of difference between modularity of the resultant and the original network across all the datasets.}
\begin{tabular}{|c|c|c|c|}
\hline
\textbf{Network} & \textbf{Link Density} & \textbf{Conductance} & \textbf{Compactness} \\ \hline
Karate           & 0.04194               & 0.00116              & 0.02219              \\ \hline
Football         & 0.04202               & 0.02193              & 0.00490              \\ \hline
Railway          & 0.06422               & 0.03174              & 0.03749              \\ \hline
Coauthorship     & 0.13037               & 0.03609              & 0.00285              \\ \hline
Amazon           & 0.13052               & 0.00550              & 0.01783              \\ \hline
Live Journal     & 0.05289               & 0.00016              & 0.03749              \\ \hline
\end{tabular}
\label{modularity_community_compare}
\end{table}

\subsection{Normalized Mutual Information}

\subsubsection{Exhaustive approach:}

Table \ref{exhaustive_nmi} presents the results of Algorithm \hyperref[algo_1]{1} on three small scale datasets where the value function that we are trying to minimize is NMI. Note that we would want to minimize NMI as this metric gives a value of 1 for two similar community structures and 0 otherwise as mentioned in Section \ref{preliminaries}. For this experiment, we fix the number of target nodes, i.e., $k = 5$. For the football network, we observe that nodes (32, 33, 5, 6, 1) are identified as the most vulnerable as they minimize the NMI score between the original and the structurally perturbed one to 0.38. This value represents the ground truth as no other combination of the five-tuple nodes will further decrease the NMI score between the two partitions. Similarly, the other small datasets' ground truth values can be found in Table \ref{exhaustive_nmi}.

% \begin{table}[ht]
\begin{table}
\centering
\captionsetup{justification=centering, font=small, skip=10pt}
\caption{Effect of the exhaustive algorithm on the smaller networks. The nodes here indicate the ID of the most vulnerable points in the network when NMI is utilized as the value function. Since the networks are smaller in size the budget $k$ was fixed at 5 which is why there the algorithm detected 5 vulnerable nodes. The corresponding NMI score reported was the minimum across all possible combinations of the nodes.}
\begin{tabular}{|c|c|c|}
\hline
\textbf{Network} & \textbf{Nodes}        & \textbf{NMI} \\ \hline
Karate           & (33, 10, 32, 6, 23)   & 0.36762      \\ \hline
Football         & (32, 33, 5, 6, 1)     & 0.38580      \\ \hline
Railway          & (51, 143, 2, 89, 287) & 0.38723      \\ \hline
\end{tabular}
\label{exhaustive_nmi}
\end{table}

\subsubsection{Network Based Greedy Approach:}

\begin{figure}
\centering
\includegraphics[width=\linewidth]{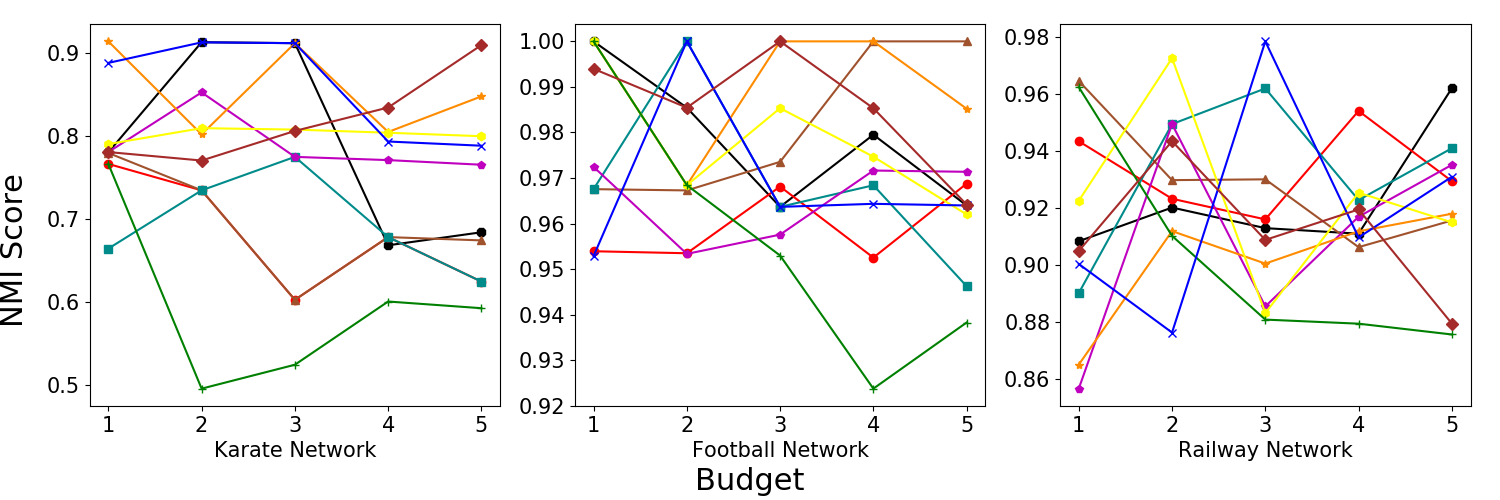}
\includegraphics[width=\linewidth]{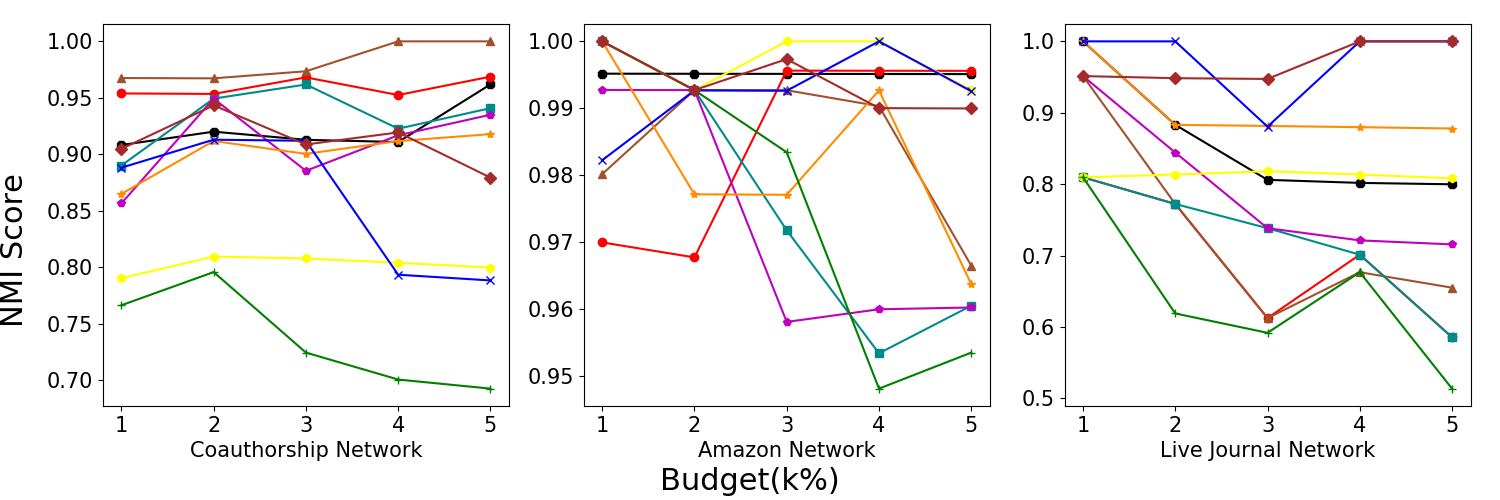}
\includegraphics[width=10cm,height=2cm,keepaspectratio]{legend.png}
\captionsetup{justification=centering, font=small}
\captionof{figure}{Results of the network-based approach over all the datasets with NMI being the target value function. For smaller networks we use $k$ = 5 and for larger ones we use 5\% of the total nodes within the network. The plots indicate that for larger budgets, eccentricity performs well in identifying vulnerable nodes when the vulnerability of a community is quantified using NMI where lower values are better indicators of disjointness. However,
upon closer inspection one can observe that these values when compared to the ground truth values reported in Table \ref{exhaustive_nmi}, are still pretty far from optimal.}
\label{nmi_network}
\end{figure}

This section presents the analysis results on all the datasets of Algorithm \hyperref[algo_2]{2}. Moreover, we fix $k = 5$ for smaller networks, as mentioned previously. Still, for more extensive networks, such removal strategy won't showcase significant effects, and hence we remove till 5\% of the total nodes in such cases. Based on Figure \ref{nmi_network}, we infer that eccentricity as a network-based greedy metric performs better than other greedy metrics when we remove the target five nodes. As we evaluate the NMI measure, we compare the minimum values attained in the ground truth data to the minimum values obtained with Algorithm \hyperref[algo_2]{2}. This is because NMI's value is small when two clusterings are not the same as mentioned previously in Section \ref{preliminaries}. Based on this comparison for smaller networks, we see that this algorithm could not attain the optimal answer indicated by Table \ref{exhaustive_nmi}. For example, in the Karate network, the minimum score obtained by Algorithm \hyperref[algo_2]{2} is  0.55, whereas the optimal answer is  0.36. This indicates that there is a lot of scope for improvement.

% \begin{figure}[!t]

\subsubsection{Community Based Greedy Approach:}

\begin{figure}
\centering
\includegraphics[width=\linewidth]{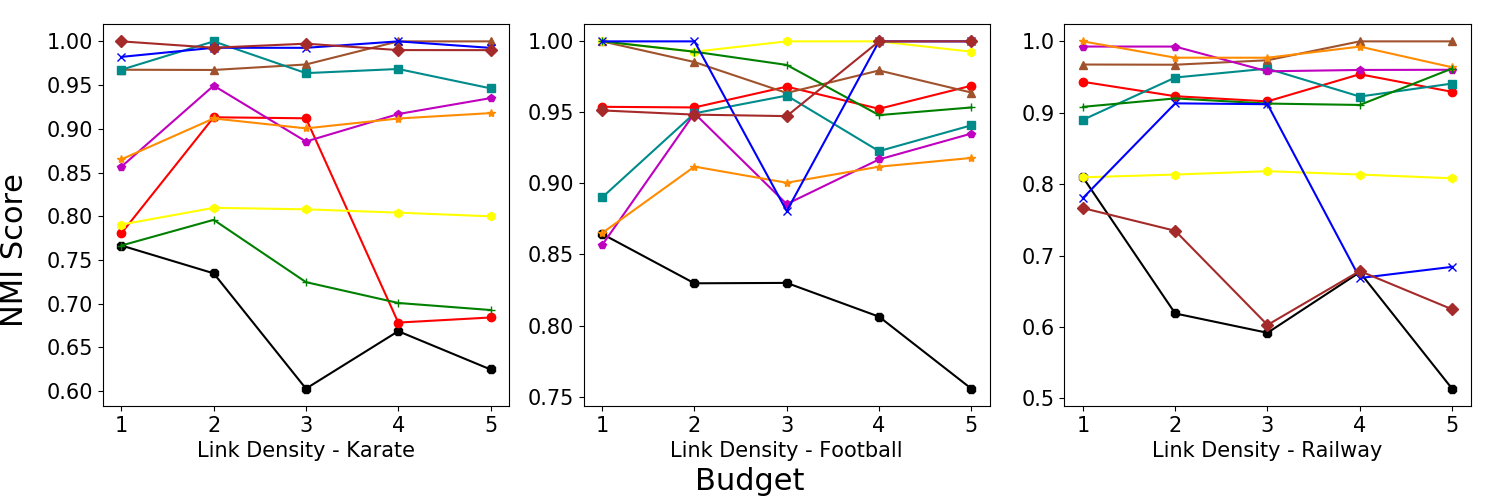}
\includegraphics[width=\linewidth]{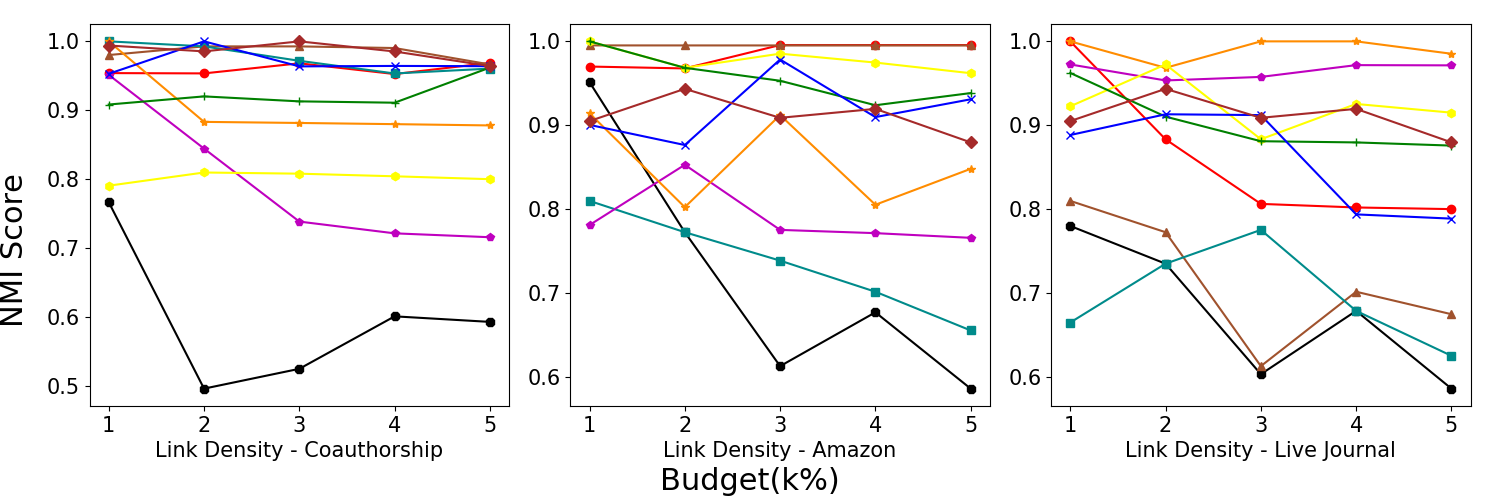}
\includegraphics[width=10cm,height=2cm,keepaspectratio]{legend.png}
\captionsetup{justification=centering, font=small}
\captionof{figure}{Outcome of the community based approach over all the datasets with NMI being the target value function. Based on Table \ref{nmi_community_compare} we observed that Link Density performs better in comparison to other greedy metric. The plots reported in this figure present the results of the node centric properties with Link Density as the community centric method. They indicate that across all the datasets clustering coefficient
performed better compared to other greedy metrics.}
\label{nmi_community}
\end{figure}

In this section, we evaluate the performance of Algorithm \hyperref[algo_3]{3} over all the datasets. As mentioned previously, we fix $k = 5$ for smaller networks and 5\% for more extensive networks. We compare the different community-centric properties in Table \ref{nmi_community_compare}. Here we present the best NMI scores obtained after applying this algorithm on all the datasets. Based on Table \ref{nmi_community_compare}, we observe that Link Density performed better than the other community-centric properties as the scores over all the datasets were minimal. With link density as the best community-centric method, we present the node centric properties' results in Figure \ref{nmi_community}. Overall the datasets we ran experiments on, we found that the clustering coefficient performs better than other greedy metrics.

Additionally, when we compare this algorithm's results with the ground truth data presented in Table \ref{exhaustive_nmi}, we observe that this solution comes close to the optimal solution. For example, in the Railway network, we follow that the best NMI score obtained to be around 0.5 is close to the ground truth score of 0.38 compared to the 0.88 score obtained from Algorithm \hyperref[algo_1]{1}. So it is evident from this data that the difference between the optimal solution and the current solution has decreased, thereby establishing the superiority of Algorithm \hyperref[algo_3]{3} over \hyperref[algo_2]{2}.

% \begin{figure}[!t]

% \begin{table}[ht]
\begin{table}
\centering
\captionsetup{justification=centering, font=small, skip=10pt}
\caption{Results of the community based approach using NMI as the target value function. It shows the effects of different community based metrics that are  used to greedily select nodes. We observe that Link Density performs better to indicate the vulnerability of nodes in terms of the NMI between the resultant and the original network across all the datasets.}
\begin{tabular}{|c|c|c|c|}
\hline
\textbf{Network} & \textbf{Link Density} & \textbf{Conductance} & \textbf{Compactness} \\ \hline
Karate           & 0.62484               & 0.68425              & 0.79993              \\ \hline
Football         & 0.75558               & 0.96877              & 0.91794              \\ \hline
Railway          & 0.51372               & 0.80825              & 0.62484              \\ \hline
Coauthorship     & 0.59279               & 0.71568              & 0.79993              \\ \hline
Amazon           & 0.58566               & 0.76560              & 0.65510              \\ \hline
Live Journal     & 0.58566               & 0.62484              & 0.78850              \\ \hline
\end{tabular}
\label{nmi_community_compare}
\end{table}

\subsection{Adjusted Rand Index}

\subsubsection{Exhaustive approach:}

Table \ref{exhaustive_ari} shows the results of Algorithm \hyperref[algo_1]{1} on three small scale datasets when using the ARI as the target value function. We performed the analysis by fixing $k = 5$. We observe that nodes (61, 85, 16, 99, 7) are the most vulnerable for the football network as their removal minimized the ARI scores between the original and the perturbed network's vertex clusterings. Similarly, the most susceptible nodes identified for the other two datasets have been tabulated in Table \ref{exhaustive_ari}. 

% \begin{table}[ht]
\begin{table}
\centering
\captionsetup{justification=centering, font=small, skip=10pt}
\caption{Effect of the exhaustive algorithm on the smaller networks. The nodes here indicate the ID of the most vulnerable points in the network when ARI is utilized as the value function. Since the networks are smaller in size the budget $k$ was fixed at 5 which is why there the algorithm detected 5 vulnerable nodes. The corresponding ARI score reported was the minimum across all possible combinations of the nodes.}
\begin{tabular}{|c|c|c|}
\hline
\textbf{Network} & \textbf{Nodes}           & \textbf{ARI} \\ \hline
Karate           & (32, 7, 12, 18, 2)       & -0.46342     \\ \hline
Football         & (61, 85, 16, 99, 7)      & 0.36342      \\ \hline
Railway          & (171, 229, 236, 75, 204) & -0.28694     \\ \hline
\end{tabular}
\label{exhaustive_ari}
\end{table}

\subsubsection{Network Based Greedy Approach:}

\begin{figure}
\centering
\includegraphics[width=\linewidth]{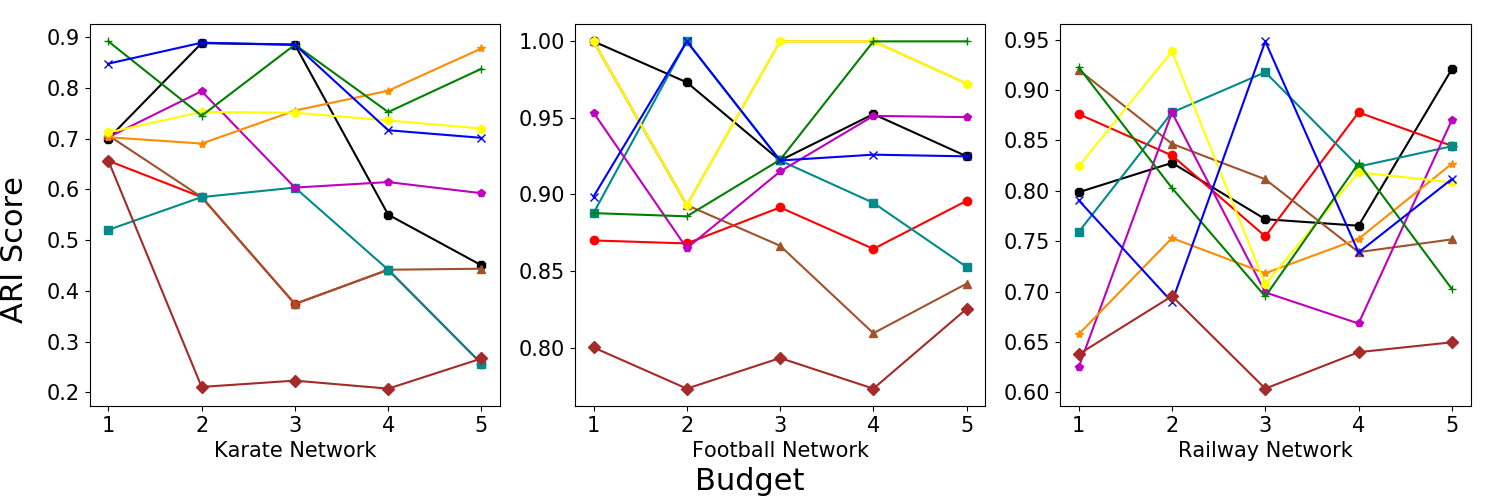}
\includegraphics[width=\linewidth]{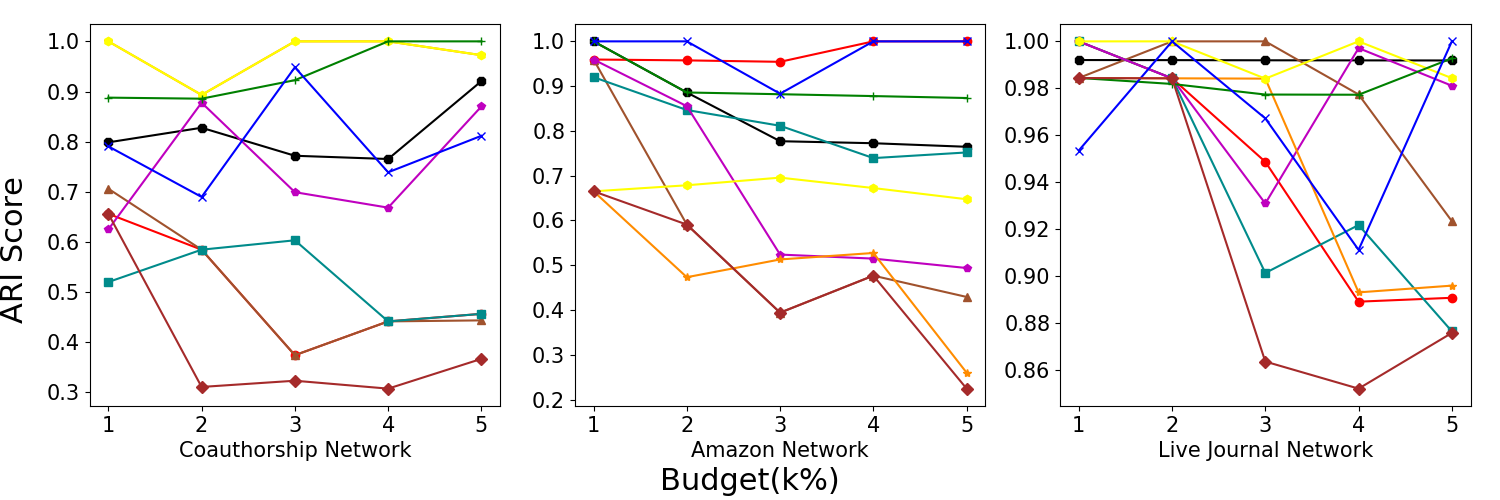}
\includegraphics[width=10cm,height=2cm,keepaspectratio]{legend.png}
\captionsetup{justification=centering, font=small}
\captionof{figure}{Results of the network based approach applied on several datasets with ARI being the target value function. We chose $k$ = 5 for smaller networks and for larger networks we chose upto 5\% of the total nodes  within the network. The plots reported in this figure show that closeness vitality performed better than the other node centric properties as for larger budgets the ARI between the resultant and the original network was low. When compared to the
exhaustive results as reported in Table \ref{exhaustive_ari}, we observe that the ARI values with closeness vitality as the greedy metric does not come close to the optimal answer.}
\label{ari_network}
\end{figure}

This section presents the analysis results on all the datasets of Algorithm \hyperref[algo_2]{2}. We fix $k = 5$ for smaller networks as mentioned previously, but for more extensive networks, we remove till 5\% of the total nodes. Based on Figure \ref{ari_network}, we infer that closeness vitality as a network-based greedy metric performs better than other greedy metrics when we remove the target five nodes. As we evaluate the ARI measure, we compare the minimum values attained in the ground truth data to the minimum values obtained with Algorithm \hyperref[algo_2]{2}. This is because ARI's value is small when two clusterings do not agree with each other, as mentioned previously in Section \ref{preliminaries}. Based on this comparison for smaller networks, we see that this algorithm cannot attain the optimal answer mentioned in Table \ref{exhaustive_ari}. For example, in the Railway network, the minimum score obtained by Algorithm \hyperref[algo_2]{2} is  0.65, whereas the optimal answer is  -0.28. This indicates that there is a lot of scope for improvement.

% \begin{figure}[!t]

\subsubsection{Community Based Greedy Approach:}

\begin{figure}
\centering
\includegraphics[width=\linewidth]{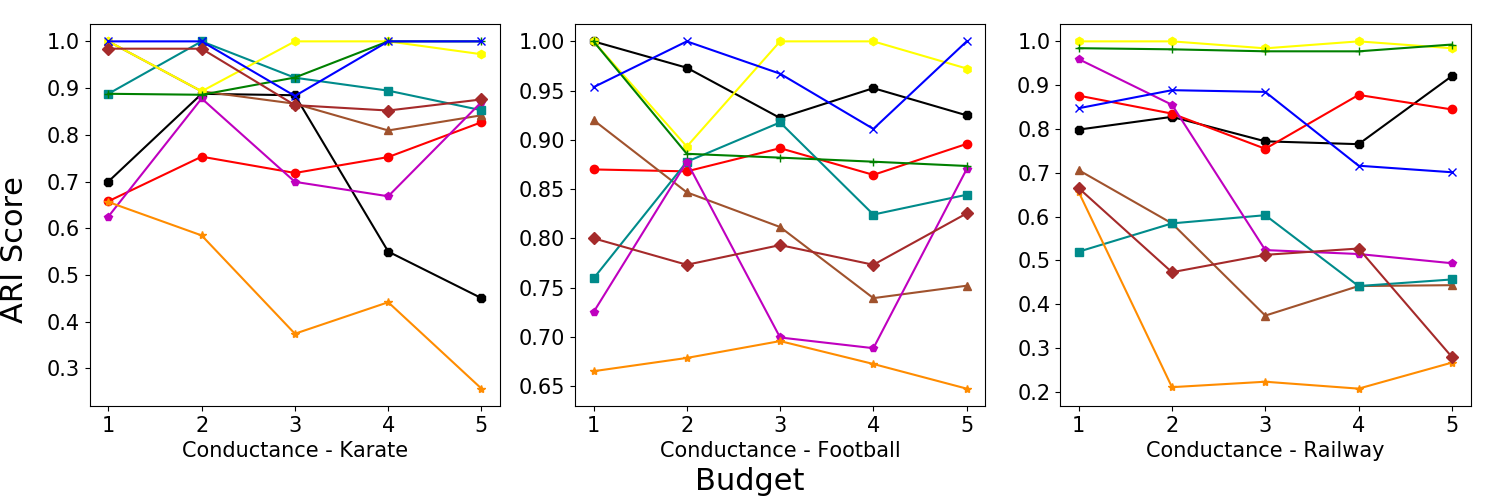}
\includegraphics[width=\linewidth]{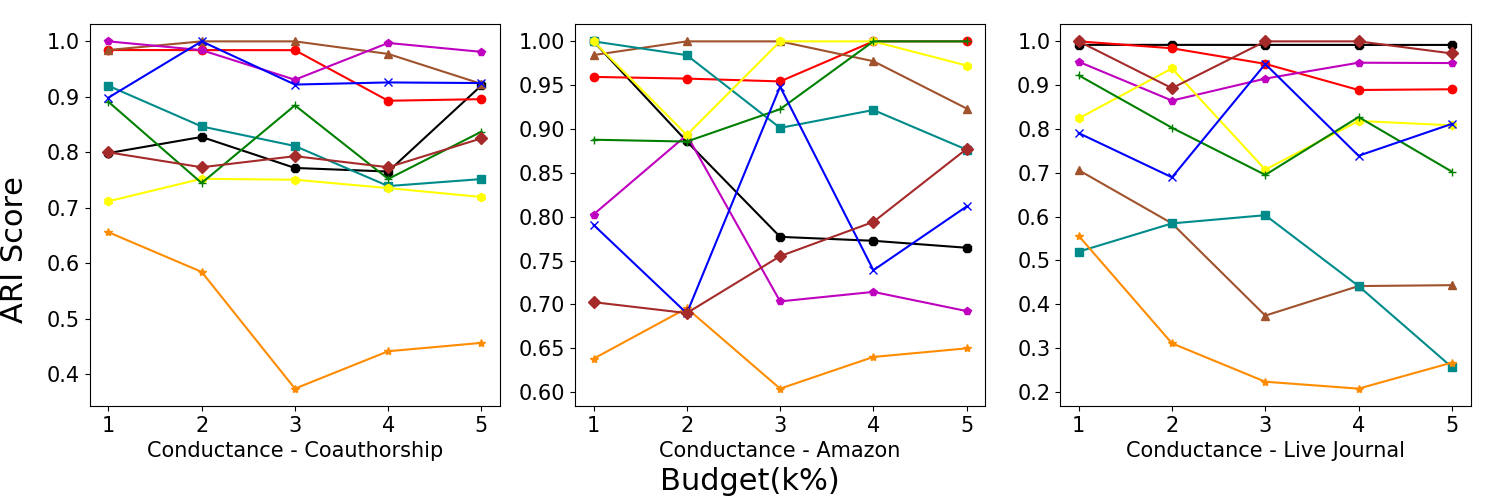}
\includegraphics[width=10cm,height=2cm,keepaspectratio]{legend.png}
\captionsetup{justification=centering, font=small}
\captionof{figure}{Outcome of the community based approach over all the datasets with ARI being the target value function. Based on Table \ref{ari_community_compare}, we observe that conductance performs better compared to other community based methods to quantify the vulnerability of communities using ARI. The plots show the effects of different node-centric properties with conductance in Algorithm \ref{algo_3}.
The results show that across all the datasets, coreness outperforms other metrics. Upon comparing these results with the ground truth data in Table \ref{exhaustive_ari}, we observe that the values are close to the optimal answers.}
\label{ari_community}
\end{figure}

In this section, we evaluate the performance of Algorithm \hyperref[algo_3]{3} over all the datasets. As mentioned previously, we fix $k = 5$ for smaller networks and 5\% for larger networks. We compare different community-centric properties in Table \ref{ari_community_compare}. Here we present the best ARI scores obtained after applying this algorithm on all the datasets. We observe that conductance performs better than the other community-centric properties as the scores over all the datasets are minimum. With conductance as the best community-centric method, we present the node-centric properties' results in Figure \ref{ari_community}. Overall the datasets we run experiments on, we find that coreness performs better compared to other metrics.

Additionally, when we compare this algorithm's results with the ground truth data presented in Table \ref{exhaustive_ari}, we observe that this solution comes close to the optimal solution. For example, in the Railway network, we follow that the best ARI score obtained to be around 0.26 is close to the ground truth score of -0.28 compared to the 0.65 score obtained from Algorithm \hyperref[algo_1]{1}. So it is evident from this data that the difference between the optimal solution and the current solution has decreased, thereby establishing the superiority of Algorithm \hyperref[algo_3]{3} over Algorithm \hyperref[algo_2]{2}.

% \begin{figure}[!t]

% \begin{table}[ht]
\begin{table}
\centering
\captionsetup{justification=centering, font=small, skip=10pt}
\caption{Results of the community based approach using ARI as the target value function. It shows the effects of different community based metrics used to greedily select nodes. We observe that conductance performs better to indicate the vulnerability of nodes in terms of the ARI between the resultant and the original community structure across all the datasets.}
\begin{tabular}{|c|c|c|c|}
\hline
\textbf{Network} & \textbf{Link Density} & \textbf{Conductance} & \textbf{Compactness} \\ \hline
Karate           & 0.45034               & 0.25670              & 0.82691              \\ \hline
Football         & 0.82530               & 0.64736              & 0.89587              \\ \hline
Railway          & 0.44367               & 0.26693              & 0.27997              \\ \hline
Coauthorship     & 0.71958               & 0.45670              & 0.75187              \\ \hline
Amazon           & 0.69230               & 0.64979              & 0.76453              \\ \hline
Live Journal     & 0.44367               & 0.25670              & 0.26693              \\ \hline
\end{tabular}
\label{ari_community_compare}
\end{table}

\subsection{Task Based Approach}

Based on the results that we observed in the previous sections for the smaller networks, we perform similar tests on more extensive networks using Algorithm \hyperref[algo_4]{4}. To quantify this algorithm's performance, we use the widely use F1 score for the link prediction task. We evaluate the fraction of active nodes at the end of the few cascades for the information diffusion task. For each experiment, we consider $k$ to be the percentage of nodes removed as otherwise the change in the community structure would not be enough to have significant effects. We have divided this section into two subsections to cover both the tasks that were described before.

\subsubsection{Link Prediction:}
We test this task by assigning probabilities to the edges using three metrics separately: Within-Inter Cluster, Modified Common Neighbors, and Modified Resource Allocation. We find that Within-Inter Cluster produces better results compared to the other alternatives.

\begin{figure}
\centering
\includegraphics[width=\linewidth]{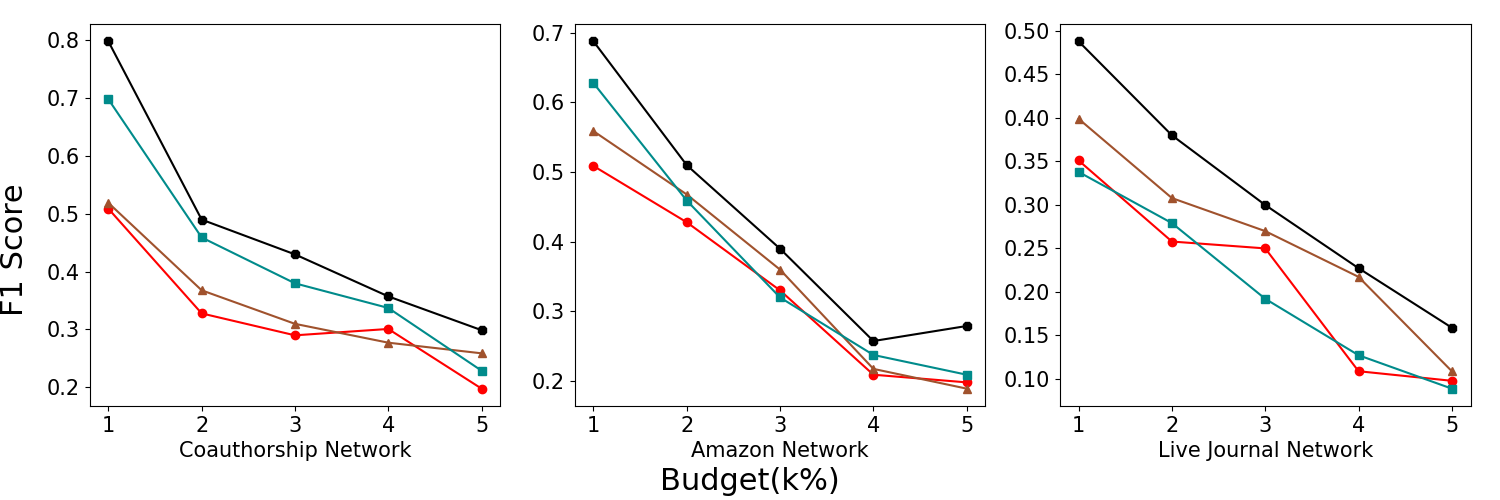}
\includegraphics[width=10cm,height=2cm,keepaspectratio]{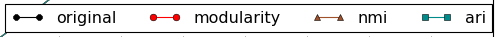}
\captionsetup{justification=centering, font=small}
\captionof{figure}{Results of link prediction task over larger datasets with all the value functions. For modularity, we use Link Density combined with eigenvector centrality. For NMI, we use Link Density combined with eccentricity, and for ARI, we use conductance and closeness vitality. Original here represents when we use the original community structure for the nodes rather than the community structure after perturbing it. The plots indicate that when using these combinations for different value functions, the link prediction task's performance quantified with the F1 score decreases.}
\label{link_prediction}
\end{figure}

Based on Figure \ref{link_prediction}, we observe that overall value functions the network's performance in the link prediction task has decreased, which is evident from the lower F1 scores. For each value function, we show the best combination as identified in the previous sections. The performance drop can be attributed to significant changes introduced into the system by removing vulnerable nodes. Their removal triggers significant structural perturbations in the underlying community structure, which causes the within-inter cluster method to assign lower probabilities to the edges due to fewer connections within the community and more connections across other communities. This decreased the likelihood of the test edge being classified as a valid link, thereby reducing the performance. 

\subsubsection{Information Diffusion:}

\begin{figure}
\centering
\includegraphics[width=\linewidth]{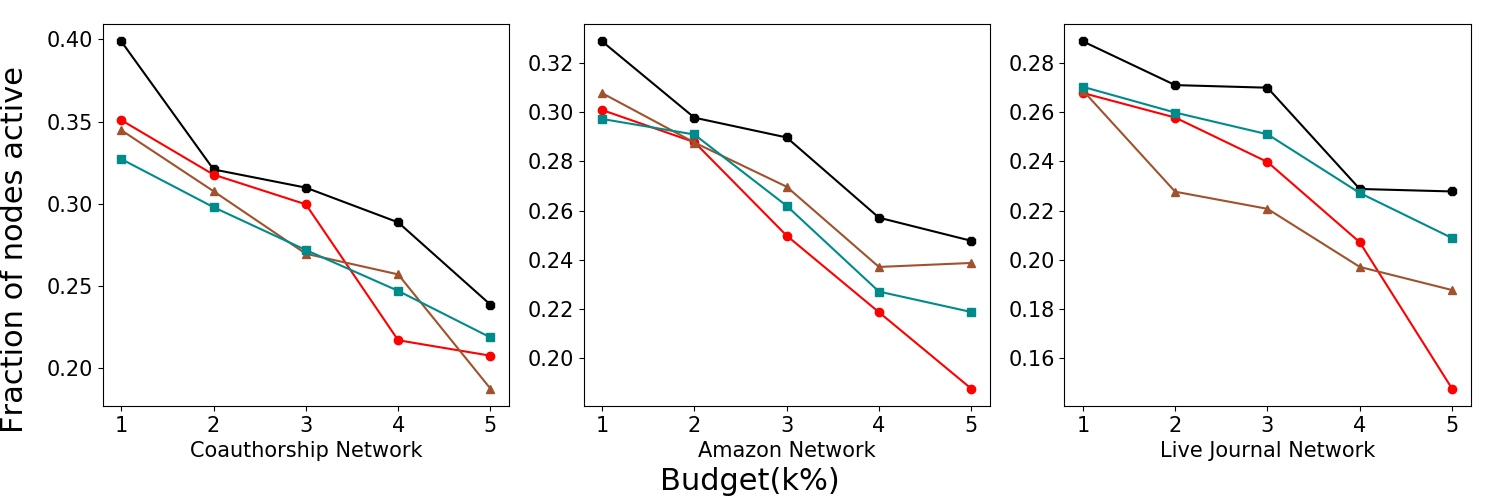}
\includegraphics[width=10cm,height=2cm,keepaspectratio]{legend_new.png}
\captionsetup{justification=centering, font=small}
\captionof{figure}{Outcome of information diffusion task over larger networks with all the value functions. For modularity we use Link Density along with eigenvector centrality, for NMI we use Link Prediction combined with eccentricity and for ARI we utilize conductance combined with closeness vitality. Original here represents when we use the original community structure for the nodes rather than the community structure after perturbing it. The plots show that for all the target value functions the performance of the information diffusion task decreases. This performance was
quantified using the fraction of \emph{active} nodes after all the iterations.}
\label{information_diffusion}
\end{figure}

In \ref{information_diffusion}, we observe that overall value functions the performance in the information diffusion task has decreased, which is evident from the lower fraction of active nodes. For this set of experiments, we set $p_i \ge p_o$ and let the cascade model run for 200 iterations. With a higher probability for the initial set and the subsequent set of active nodes to affect the nodes within their community, it is trivial to see that the fraction of nodes that will be active at the end of all the iterations would be low. This is true if the underlying community structure was significantly perturbed and the network was highly disconnected, whereas it would be the opposite for the other case.  For each value function, we show the best combination as identified in the previous sections.

This shows that the best combination of community-centric and network-centric nodes that we get from Algorithm \hyperref[algo_3]{3} when applied to the more extensive networks using Algorithm \hyperref[algo_4]{4} results in the decrease in the performance of the networks over both tasks that they are employed on, thereby validating our initial hypothesis. This establishes that Algorithm \hyperref[algo_3]{3} can be applied to any general network irrespective of the size.

% \begin{figure}[!t]

\section{Conclusion}
\label{conclusion}

In this paper, we proposed a hierarchical greedy-based approach that efficiently identified critical nodes in the network, which significantly impacted the underlying community structure. Additionally, we also proposed a novel task-based strategy to apply the results of the hierarchical greedy based approach on more extensive networks and quantify its effectiveness, which would enable us to estimate the performance of the algorithm in a real-world context. 

Due to the extensive size of our experiments, we show our best results only. Since Algorithm \hyperref[algo_1]{1} is exhaustive and hence was applied only to small networks such as Karate, Football and Railway Network. The results of this algorithm provided us with the benchmark to compare with our other algorithms. We further saw that Algorithm \hyperref[algo_2]{2} was not that promising and were far from the gold standard in comparison to Algorithm \hyperref[algo_3]{3} which came close to the gold standard. This comparison showed that Algorithm \ref{algo_3} works best for small networks. As mentioned previously, we used Algorithm \hyperref[algo_4]{4} to compare the performance of Algorithm \hyperref[algo_2]{2} and Algorithm \hyperref[algo_3]{3} for large networks such as Co-authorship, Amazon, and Live Journal Networks. Based on these results, we established that when we use Algorithm \hyperref[algo_3]{3}, we get a performance drop over both the tasks, namely link prediction and information diffusion, compared to the original network. This establishes the generalizability of Algorithm \hyperref[algo_3]{3}.

To conclude, this work has provided a hierarchical approach that allowed for identifying the vulnerable nodes in a network efficiently. The proposed method was used to analyze the community vulnerability of several networks whose validity was established using both exhaustive and task-based approaches depending on the network's size.

\if 0
As part of the future work, we will attempt to provide a more mathematical construct to explain these empirical results. With many community detection algorithms around, we will try to expand the proposed algorithm results to them and report a more comprehensive overview of those results. We leave this line of research as part of the future plan.  
\fi

\subsubsection*{Acknowledgement.}
T. Chakraborty would like to acknowledge the support of  SERB (ECR/2017/001691) and the Infosys Centre for AI, IIIT-Delhi.

\bibliographystyle{splncs04}
\bibliography{LNCSbib}
\end{document}